# Ozone adsorption on graphene: ab initio study and experimental validation

*Geunsik Lee,[b] Bongki Lee,[a] Jiyoung Kim,[a] and Kyeongjae Cho*[a,b]*

[a]Department of Materials Science and Engineering and [b]Department of Physics, The University of Texas at Dallas, Richardson, Texas 75080, USA

TITLE RUNNING HEAD: Ozone on graphene: ab initio study and experiment

*Corresponding author: kjcho@utdallas.edu

We have investigated ozone adsorption on graphene using the ab initio density functional theory method. Ozone molecules adsorb on graphene basal plane with binding energy of 0.25 eV, and the physisorbed molecule can chemically react with graphene to form an epoxide group and an oxygen molecule. The activation energy barrier from physisorption to chemisorption is 0.72 eV, and the chemisorbed state has the binding energy of 0.33 eV. These binding energies and energy barrier indicate that the ozone adsorption on graphene is gentle and reversible. Atomic layer deposition experiment on ozone treated graphite has confirmed the presence of uniform hydrophilic groups on graphene basal plane. This finding can be applied to diverse chemical functionalization of graphene basal planes.

Graphene is a planar form of sp$^2$-bonded carbon materials and has emerged as a promising nanoelectronic device material along with related carbon nanotubes (CNTs) and fullerenes.[1-3] The interaction between sp$^2$-bonded carbon materials and molecular species has been extensively investigated for carbon nanotubes and fullerenes for sensor applications and chemical functionalization.[4-7] A detailed theoretical study has shown that the reactivity of sp$^2$-bonded carbon materials strongly depends on the local curvature of the surface leading to high reactivity of fullerenes followed by nanotubes and weakly reactive graphene.[7-9] Recent experimental studies have shown that the graphene basal planes are very inert against chemical interaction with molecular species.[10-12] Furthermore, oxygen molecule is known to adsorb on nanotube (CNT) surface and induce chemical doping[13] whereas graphene is inert against oxygen adsorption.[14] For practical device applications of graphene, experimental groups have developed diverse non-covalent[12,15-17] or covalent[18] approaches to functionalize the graphene basal plane. Recently, Kim et al. have shown that ozone can provide a facile route to functionalize basal plane of highly oriented pyrolytic graphite (HOPG) leading to uniform growth of Al$_2$O$_3$ by atomic layer deposition (ALD).[11]

Ozone interaction with carbon nanotubes has been studied experimentally[19-21] and theoretically.[22,23] Experimental studies have shown that ozone-reacted CNTs (SWCNTs) have surface-bound functional groups such as C=O, C=C and C-O and a heat treatment (T > 473 K) of the ozone-reacted CNT is shown to release gas phase CO$_2$ and CO.[19] It is also shown that ozone exposure induces p-type doping and correspondingly reduces resistance in CNTs.[20,21] Theoretical studies have shown that the binding energy of ozone on CNT is 0.2~0.3 eV depending on the diameter and chirality.[20,22-24] Since the physisorption energy was too small to explain ozone exposure effect on the CNT resistance, it was speculated that some defects sites would be responsible for the ozone adsorption and the corresponding resistance change.[20] These studies indicate that the understanding on the ozone interaction with CNT is not well established, yet. According to the theoretical study of Park et al.,[9] surface reactivity of graphene would be smaller than that of CNTs, and ozone is expected to interact weakly with graphene.

A weak binding (< 0.3 eV) of ozone on graphene is not consistent with the experimental demonstration of an ozone treatment effect on HOPG.[11] To understand the underlying mechanisms of ozone interaction with graphene, it is necessary to investigate diverse reaction pathways of ozone adsorption on graphene. In this paper, we have performed a set of systematic ab initio density functional theory (DFT) simulations to examine the ozone molecule interaction with graphene surface and identified physisorption and chemisorption binding configurations and the transition states connecting them. The calculated binding energies and activation barrier are tested by a systematic experimental study of controlled ozone exposure on graphene followed by ALD growth of alumina.

In order to calculate the total energy of ozone adsorption on graphene, we have used Vienna ab initio Simulation Package (VASP)[25] with projection augmented wave (PAW) pseudopotentials[26] within the local spin density approximation (LSDA).[27] LSDA rather than LDA (local density approximation) is necessary to properly model the ozone chemisorption on graphene leading to oxygen molecule formation.[28] We have employed a plane wave basis set with an energy cutoff of 400 eV which is shown to give a good convergence. We have used a supercell of 3×3 periodic unit cells of graphene to minimize the interaction between periodic images. The resulting image distance is 7.38 Å which may not be large enough to completely remove the image interactions. However, the image interaction energy remains relatively constant during ozone adsorption on graphene leading to negligible effect on the calculated binding energies. The vertical unit cell size was chosen to be 15 Å. The k-point grid of 5×5×1 was adopted for the Brillouin zone sampling. One ozone molecule was placed above a graphene plane and the distance between the molecule and the plane was reduced from 5 Å to calculate the binding energy curves at different surface binding sites. As the ozone molecule was approaching the graphene plane, atomic structure was fully relaxed at each height of ozone relative to graphene sheet. The atomic positions were optimized until the total energy changes less than 0.001 eV under the constraint that two C atoms (indicated by gray circles in Figure 1a) were fixed to keep the relative height of ozone molecule to the graphene plane. When ozone dissociative chemisorption occurs on graphene with a finite energy barrier, we additionally performed climbing nudged elastic band (NEB) calculation to estimate accurate energy barrier.[29] This method allows us to find the saddle point with a small number of images along the reaction path. Figure 1a shows the 3×3 graphene supercell with three adsorption (top, center, and bridge) sites indicated, and Figure 1b shows 12 different ozone molecule adsorption configurations indexed as A – L with the corresponding binding energy curves shown in Figure 2.

Our experiments were performed using the following method. An atomically smooth surface with sharp step edges was prepared by peeling off several top layers of highly oriented pyrolytic graphite (HOPG) using an adhesive tape. Immediately after the mechanical exfoliation, the HOPG samples were transferred into a commercial ALD reactor (Cambridge Nanotech Inc., Savannah 100). Prior to $Al_2O_3$ deposition, the HOPG surface was treated by ozone ($O_3$) gas (22 wt%, In-USA) with different ozone exposure times (10, 30, and 60 sec) at a substrate temperature of 500 K in the ALD reactor. Following the ozone exposure, 20 cycles of $Al_2O_3$ deposition were carried out at a substrate temperature of 500 K using tri-methyl aluminum (TMA) and water ($H_2O$) as reactants, respectively. A more detailed description of experimental procedure can be found elsewhere.[11] The deposited $Al_2O_3$ layers onto ozone-treated HOPG surface were characterized by Veeco multimode AFM in tapping mode.

Figure 2 shows the binding energy curves of ozone on graphene at different adsorption sites and molecule configurations relative to graphene plane (12 geometries shown in Figure 1b A – L). The height (h) of ozone molecule is defined as the vertical distance between the lowest O atom of the ozone molecule and the fixed carbon atoms in graphene plane as shown in Figure 1b. We have examined four different ozone molecule configurations as it approaches the graphene plane from h = 5 Å: (i) V shape (shown in Figure 1b A – E), (ii) Λ shape (shown in Figure 1b F – H), (iii) planar (shown in Figure 1b I – J) and (iv) vertical (shown in Figure 1b K – L) $O_3$ configurations. For V shape ozone configuration, three adsorption sites (the lowest O atom on top, center, and bridge sites) are examined in the DFT calculations, and the binding energy curves are shown in Figure 2a. The binding energy curves show weak physisorption states with binding distance of h = 2.8 Å and adsorption energies of 0.11, 0.11 and 0.15 eV for top (A), bridge (B), and center (C) sites, respectively. To further examine the effect of $O_3$ configuration on the binding energy curve, ozone geometries at top and center sites are rotated around the vertical axis as shown in Figure 1b D and E. Figure 2a shows that such rotation does not change the binding energy curves. All five binding energy curves monotonically increase as h decreases from 2.8 Å and indicate no chemisorption for these binding configurations. This result is consistent with the stability of middle oxygen atom in O-O-O structure since it is fully saturated by two neighboring O atoms.

For Λ shape ozone configuration, Figure 2b shows three binding energy curves, and the lowest physisorption energy is 0.24 eV at h = 2.8 Å at the center adsorption site (assigned by two O atoms at the bottom of Λ shape). For the center site (curve H), the binding energy curve monotonically increases as h decreases similar to V shape $O_3$ in Figure 2a. However, at the top and bridge sites (curves F and H in Figure 2b), the ozone molecule has an additional chemisorption state at h = 1.5 Å with the adsorption energy of 0.26 eV. The energy barrier from the physisorption to the chemisorption state is lower for the top (F) than the bridge (G) in Figure 2b. The minimum energy path from our NEB calculation is shown in the inset. The transition state energies are 0.53 (F), 0.65 (G) eV higher than that of the isolated ozone at h = 5.0 Å. Since their physisorption states have 0.21 eV lower energies than the isolated one, the corresponding energy barriers toward chemisorption are 0.74, 0.86 eV for the F and G, respectively.

For planar $O_3$ configuration (Figure 1b I and J), the binding energy curves on top and bridge sites are shown in Figure 2c. For vertical configuration (Figure 1b K and L), the same curves are shown in Figure 2d. Since one of the ozone terminal atoms is on top or bridge site (similar to the Λ shape), the binding energy curves show similar physisorption-chemisorption transition behaviors. The top site shows lower energy barrier than the bridge site in binding energy curves of Figures 2c and 2d, so we performed NEB calculations for the top case, which are shown in the insets. For the planar (vertical) ozone, the transition state has 0.51 (0.47) eV higher energy than the isolated one, and the physisorption and chemisorption binding energies are 0.25 (0.17) and 0.33 (0.22) eV. Thus the energy barrier of the physisorption- chemisorption transition is 0.76 (0.64) eV for the planar (vertical) case.

Although our calculated transition state energy for different ozone configurations varies a little, ozone molecules will follow the minimum energy path during the dissociative chemical adsorption. In Figure 3, we show minimum energy configurations for each of physisorption, transition and chemisorption states. As the ozone molecule goes over the energy barrier from the physisorption state (the left part), it is going through a chemical reaction of breaking one of O-O bonds and forming O-C bond at the transition state (the middle part) and a subsequent reaction of forming two new C-O bonds in the epoxide group (C-O-C of the right chemisorption state). It is clear that the ozone molecule is dissociated into an epoxide group (C-O-C) on graphene basal plane and a separate oxygen molecule. This feature is practically the same for all four chemical adsorptions shown in Figure 2. As indicated in Figure 3, our lowest physisorption and chemisorption binding energies are 0.25 and 0.33 eV. The energy barriers for the physisorption-to-chemisorption and the reverse reactions are 0.72 and 0.80 eV. In order to study reaction properties of $O_3$ and graphene in the following, we use these energy values.

We estimate the coverage of physisorbed ozone molecules by assuming thermal equilibrium between physisorbed and gas phases. Our calculated physisorption energy is 0.25 eV. The curvature of the binding energy curve at the bottom of physisorption state at h = 2.8 Å gives the force constant of 0.5 eV/Å$^2$ and the corresponding attempt frequency of $v_0 = 2 \times 10^{12}$ sec$^{-1}$.[30] By using these two quantities and the Arrhenius equation, we can compute the desorption rate as $v_0 \exp(-0.25/k_B T) \sim 10^9$ at T=500 K. The adsorption rate is related with the density of ozone arriving at the surface per unit time and its sticking probability. Note that a similar binding behavior was found for $NO_2$ on CNT surface,[6] and we use similar sticking coefficient found in ref. 7. By using Langmuir Isotherm model,[31] we estimate $10^{-7}$ ML, where 1 ML is defined as the number of molecules as surface area divided by ozone cross section, for the physisorption coverage under the same condition of ozone partial pressure (0.01 MPa) as in our experimental work (see supplementary material for details). This coverage is too low to provide effective graphene surface modification which has been shown in the uniform ALD growth of alumina on ozone treated HOPG.[11]

We suggest that the plausible mechanism should involve ozone chemisorption. From our calculation, the physisorbed ozone molecules have the activation energy of 0.72 eV to form epoxide groups on graphene and oxygen molecules. It also shows that the epoxide group has the C-O bonding distance of 1.44 Å, and that the desorption energy of an epoxide group into an atomic oxygen is 3.23 eV. Therefore, the epoxide groups are very stable in the absence of reverse reaction. The reverse reaction to the epoxide formation (i.e., transition from right to left configurations in Figure 3) is an epoxide group interaction with an oxygen molecule to form an ozone molecule. It requires ambient oxygen molecules, and the net result of the reaction is the reduction of oxidized graphene by oxygen molecules. The activation barrier of desorption is calculated to be 0.80 eV, and the desorption rate is proportional to the oxygen density arriving on the graphene surface. At the equilibrium between the dissociative chemisorptions and the reverse reactions, we can estimate the epoxide coverage as described in the supplementary material. Under the same gas pressure and temperature as in our experiment, our estimation gives the epoxide coverage of 0.5 ML (see Figure S2). In order to estimate the time for reaching the equilibrium, we compute the rate of chemisorption at experimental temperature (T=500 K). By using $v_0 \exp(-0.72/k_B T)$, we get the reaction rate of $7 \times 10^4$ sec$^{-1}$ which corresponds to the reaction time of $1.43 \times 10^{-5}$ sec. Although the physisorbed $O_3$ coverage was found to be as low as $10^{-7}$ ML, it can reach about 0.1 ML after ten seconds during which million times of physisorption-to-chemisorptions conversion occur. During this time the chemisorption is dominating over the reverse reaction. However, the rate of the reverse reaction increases as the epoxide coverage increases, and the rates of both reactions will eventually become equal resulting in the equilibrium. This reaction rate and resulting equilibrium coverage is reasonable to provide hydrophilic reaction sites for ALD growth as shown in the following experimental data.

To validate the theoretical finding of ozone-induced epoxide formation on graphene basal planes, we have performed a systematic experimental study on ozone treatment of graphite with different ozone exposure time (10, 30, and 60 seconds at 500 K) immediately followed by 20 cycles of TMA/$H_2O$ for $Al_2O_3$ depositions in the same ALD chamber (i.e. without an exposure to oxygen atmosphere). The AFM image of HOPG after ozone exposure of 10 seconds reveals that an atomically smooth surface was maintained with a RMS roughness of ~ 0.1 nm in 1.5 μm × 1.5 μm area as indicated in Figure 4a. However, the samples exposed by ozone for 30 and 60 seconds clearly show etching on the basal planes ranging from one-layer to multiple-layer deep pits with increased ozone exposure time as shown in Figures 4b and 4c. To test the presence of functional groups on the atomically smooth surface, 20 cycles of $Al_2O_3$ deposition using TMA/$H_2O$ was carried out on the sample shown in Figure 4a. We found that $Al_2O_3$ layer was deposited along with step edges as well as on graphite basal planes with a uniform nucleation of the $Al_2O_3$ film as shown in Figure 4d. Furthermore, the height of $Al_2O_3$ layer deposited onto the basal planes was found to be ~ 2 nm, corresponding to a typical growth rate of ~ 0.1 Å/cycle for TMA/ $H_2O$ process. Since graphite step edges are known to be reactive toward ALD precursors due to dangling bonds of broken C-C σ bonds, the growth of the $Al_2O_3$ film is initiated at the graphite step edges without any nucleation on the basal plane when TMA/$H_2O$ cycles are applied on graphite without ozone treatment. As for the observed etching at longer ozone exposure times, 30 and 60 seconds shown in Figures 4b and 4c, thermally induced migrations of epoxides would break C-C bonds probably releasing CO and $CO_2$ gas molecules as reported in the literatures,[32,33] but the detailed etching mechanism is not well understood yet.

Consequently, the observation of the nucleation of the $Al_2O_3$ film on the basal plane validates our theoretical prediction that ozone will form epoxide groups on graphene basal plane, and a uniformly distributed epoxide groups will interact with TMA to nucleate $Al_2O_3$ film on the graphene basal plane. The XPS spectra of ALD grown alumina on ozone treated HOPG has shown the signature of C-O-C peak in C 1s spectrum indicating that epoxide groups were present to facilitate the ALD growth.[11] Even though the sample was exposed to air before XPS measurement, the residual epoxide groups were protected by alumina film. However, the XPS data of ozone treated graphite do not show epoxide signature after air exposure, and this result is consistent with the theoretical prediction of oxygen induced epoxide removal in the absence of ozone gas. These experimental results are consistent with the theoretically predicted mechanisms of ozone interaction with graphene and validate the model study presented in this paper.

In summary, we have studied ozone interaction with graphene by calculating the total energy of ozone adsorption on graphene as a function of ozone height from the graphene basal plane. We have found physisorption states at the height of 2.8 Å for all binding configurations and dissociative chemisorption states for the top and bridge sites adsorption. The chemisorption state forms an epoxide group on graphene and a gas phase oxygen molecule. The effective energy barrier for epoxide formation from ozone gas molecule is 0.47 eV leading to rapid epoxide formation under the ozone exposure. Under oxygen exposure, epoxide groups can be desorbed with an activation energy of 0.80 eV. Experimental study of ALD growth on ozone treated graphite has confirmed that an ozone exposure to graphite creates uniformly distributed hydrophilic reaction sites on the graphene basal plane leading to a uniform nucleation of alumina film during 20 cycles of $TMA/H_2O$ reactions. The theoretical study of ozone adsorption on graphene validated by ALD experiment provides a detailed mechanistic understanding on how to functionalize graphene basal plane through a reversible epoxide group formation. Graphene with controlled epoxide functionalization can be used as a starting material for diverse chemical functionalization through a subsequent chemical modification of epoxide groups on graphene basal planes.

**Acknowledgement.**

This work is supported by NRI-SWAN project. JK and BL acknowledge financial supports of NRI-MIND and System IC (COSAR/MKE) in Korea. We thank Prof. Y. Chabal and Dr. B. Shan for helpful suggestions.

**Supporting Information Available**

Details on the following topics are available: derivation of physisorption and chemisorption coverages of ozone at the equilibrium, epoxide coverage and growth rate as functions of temperature. This information is available free of charge via the Internet at http://pubs.acs.org.

**References**

(1) Novoselov, K. S.; Geim, A. K.; Morozov, S. V.; Jiang, D.; Zhang, Y.; Dubonos, S. V.; Grigorieva, I. V.; Firsov, A. A. *Science* **2004**, *306*, 666.
(2) Zhang, Y.; Tan, Y.-W.; Stormer, H. L.; Kim, P. *Nature* **2005,** 438, 201.
(3) Geim, A. K.; Novoselov, K. S. *Nature Mater.* **2007**, *6*, 183.
(4) Kong, J.; Franklin, N.; Zhou, C.; Chapline, M. G.; Peng, S.; Cho, K.; Dai, H. *Science* **2000**, *287*, 622.
(5) Schedin, F.; Geim, A. K.; Morozov, S. V.; Hill, E. W.; Blake, P.; Katsnelson, M. I.; Novoselov, K. S. *Nature Mater.* **2007**, 6, 652.
(6) Peng, S.; Cho, K. *Nanotechnology* **2000**, *11*, 57.
(7) Peng, S.; Cho, K.; Qi, P.; Dai, H. *Chem. Phys. Lett.* **2004**, *387*, 271.
(8) Park, S.; Srivastava, D.; Cho, K. *Nanotechnology* **2001**, *12*, 245.
(9) Park, S.; Srivastava, D.; Cho, K. *Nano Lett.* **2003**, *3*, 1273.
(10) Xuan, Y.; Wu, Y. Q.; Shen, T.; Qi, M.; Capano, M. A.; Cooper, J. A.; Ye, P. D. *Appl. Phys. Lett.* **2008**, 92, 013101.
(11) Lee, B.; Park, S.-Y.; Kim, H.-C.; Cho, K.; Vogel, E. M; Kim, M. J.; Wallace, R. M.; Kim, J. *Appl. Phys. Lett.* **2008**, *92*, 203102.
(12) Wang, X.; Tabakman, S. M.; Dai, H. *J. Am. Chem. Soc.* **2008**, *130*, 8152.
(13) Jhi, S.-H.; Louie, S. G.; Cohen, M. L. *Phys. Rev. Lett.* **2000**, *85*, 1710.
(14) Yu, S. S.; Zheng, W. T.; Jiang, Q. *IEEE T. Nanotechnol.* **2008**, *7*, 628.
(15) Williams, J. R.; DiCarlo, L.; Marcus, C. M. *Science* **2007**, *317*, 638.
(16) Farmer, D.; R. Gordon, *Nano Lett.* **2006**, *6*, 699.
(17) Lu, Y.; Bangsaruntip, S.; Wang, X.; Zhang, L.; Nishi, Y.; Dai, H. *J. Am. Chem. Soc.* **2006**, *128*, 3518.
(18) Banerjee, W.; Hemraj-Benny, T.; Wong, S. S.; *Adv. Mater.* **2005**, *17*, 17.
(19) Mawhinney, D. B.; Naumenko, V.; Kuznetsova, A.; Yates, J. T.; Liu, J.; Smalley, R. E. *J. Am. Chem. Soc* **2000**, *122*, 2383.
(20) Picozzi, S.; Santucci, S.; Lozzi, L.; Cantalini, C.; Baratto, C.; Sberveglieri, G.; Armentano, I.; Kenny, J. M.; Valentini, L.; Delley, B. *J. Vac. Sci. Technol A* **2004**, *22*, 1466.
(21) Wongwiriyapan, W.; Honda, S.; Konishi, H.; Mizuta, T.; Ikuno, T.; Ohmori, T.; Ito, T.; Shimazaki, R.; Maekawa, T.; Suzuki, K.; Ishikawa, H.; Oura, K.; Katayama, M. *Jpn. J. Appl. Phys.* **2006**, *45*, 3669.
(22) Yim, W. L.; Liu, Z. F. *Chem. Phys. Lett.* **2004**, *398*, 297.
(23) Akdim, B.; Kar, T.; Duan, X.; Pachter, R. *Chem. Phys. Lett.* **2007**, *445*, 281.
(24) Picozzi, S.; Santucci, S.; Lozzi, L.; Valentini, L.; Delley, B. *J. Chem. Phys.* **2004**, *120*, 7147.
(25) Kresse, G.; Hafner, J. *Phys. Rev. B* **1993**, *47*, 558; Kresse, G.; Furthmüller, J. *ibid.* **1996**, 54, 11169.
(26) Blöchl, P. E. *Phys. Rev. B* **1994**, 50, 17953; Kresse, G.; Joubert, D. *Phys. Rev. B* **1999**, 59, 1758.
(27) Ceperley, D. M.; Alder, B. J. *Phys. Rev. Lett.* **1980**, *45*, 566.

(28) In the chemisorption reaction, $O_3 \rightarrow O(chemisorbed) + O_2$, the spin polarized $O_2$ with $2\mu_B$ magnetic moment has 0.69 eV lower energy than non-spin polarized one. So, usual LDA calculation of the same reaction pathway does not show stable chemisorption states.
(29) Henkelman, G.; Uberuaga, B. P.; Jónsson, H. *J. Chem. Phys.* **2000,** *113,* 9901.
(30) We have used harmonic potential energy ($kx^2/2$) to get the force constant of ozone molecule vibration along the surface normal direction with respect to the equilibrium physisorption state. From the fitted force constant ($k$), we obtained the frequency from $v_0 = \sqrt{k/M}$, where $M$ is the ozone molecule mass.
(31) Somorjai, G. A. *Introduction to Surface Chemistry and Catalysis*; Wiley: New York, USA, 1994; p. 303.
(32) Li, J.-L.; Kudin, K. N.; McAllister, M. J.; Prud'homme, R. K.; Aksay, I. A.; Car, R. *Phys. Rev. Lett.* **2006**, *96*, 176101.
(33) Paci, J. T.; Belytschko, T.; Schatz, G. C. *J. Phys. Chem. C* **2007**, *111*, 18099.

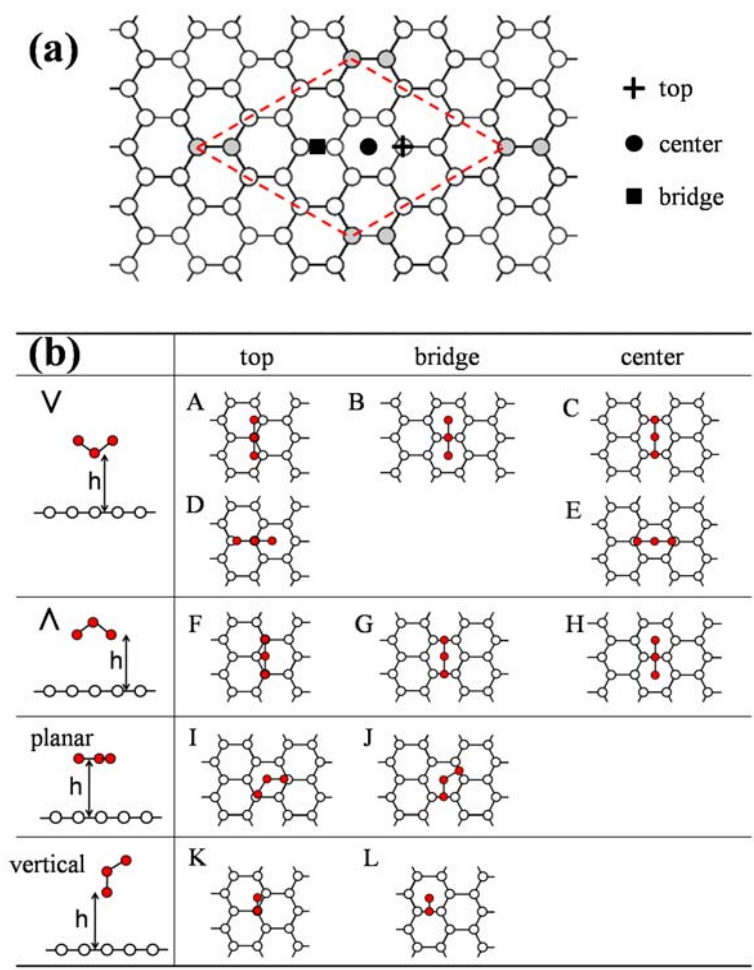

**Figure 1.** (a) 3×3 unit cell of graphene (dashed line) with two fixed carbon atoms (gray circles). Three adsorption sites for ozone molecule: top (plus), center (circle), bridge (rectangle). (b) 12 different ozone

molecule adsorption configurations indexed as A – L: V shape (A – E), Λ shape (F – H), planar (I, J), and vertical (K, L) O$_3$ configurations and three adsorption sites.

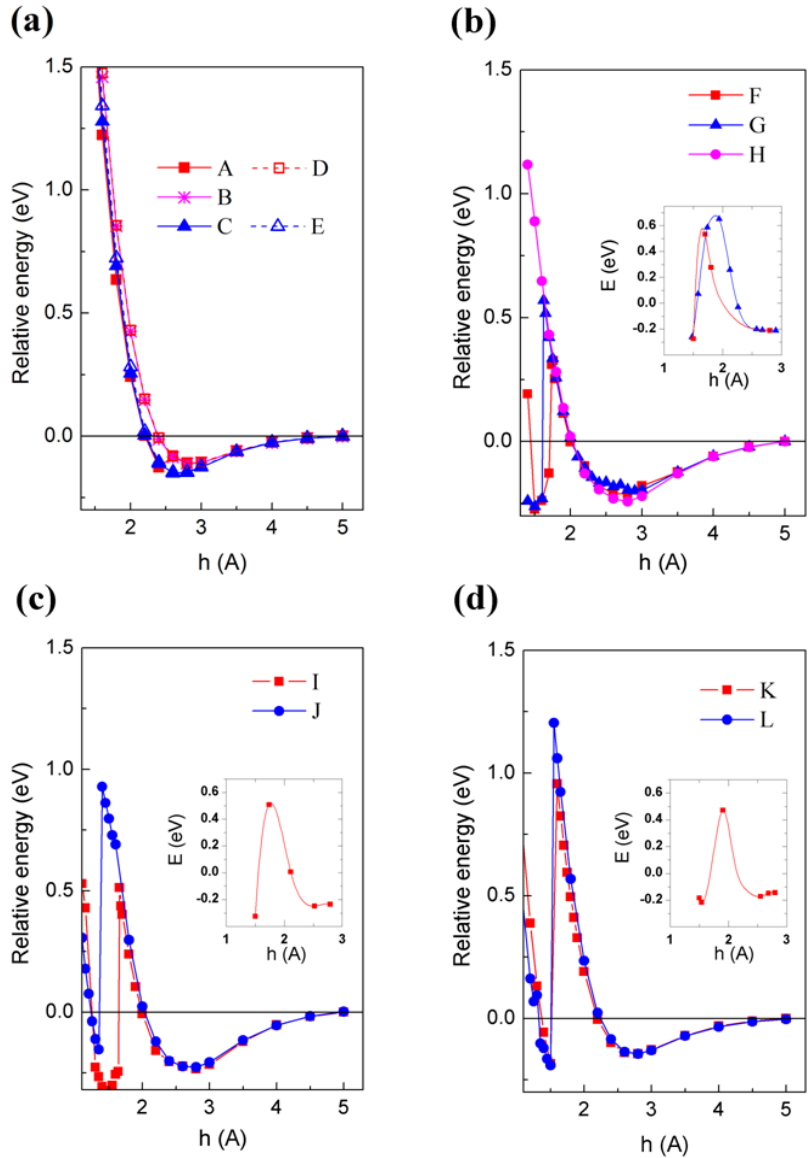

**Figure 2**. The total binding energy curves of ozone on graphene system as a function of height (h) for 12 different configurations shown in Figure 1b. (a) Binding energy curves of V shape (A – E) ozone configuration with the lowest O atom approaching to the top (A and D), bridge (B), center (C and E) sites. (b) Binding energy curves of Λ shape (F – H) ozone configuration with the two lowest O atoms approaching to the top (F), bridge (G), center (H) sites. (c) Binding energy curves of planar ozone with the both ends approaching the top sites (I) and the bridge sites (J). (d) Binding energy curves of vertical ozone with the lowest O atom approaching the top site (K) and the bridge site (L). In the insets of (b)-

(d), the NEB calculation results are shown for the F, G, I and K configurations, where the line is only for guidance of eye.

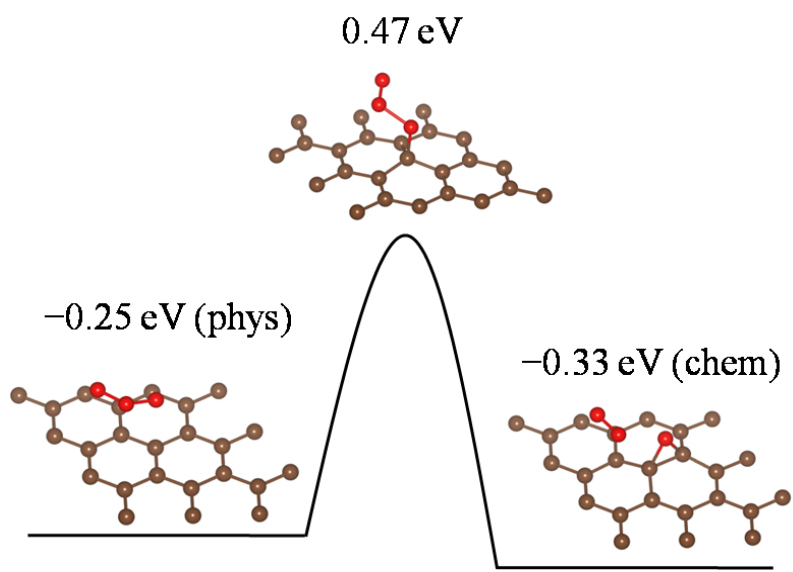

**Figure 3**. Dissociative chemisorption of an ozone molecule from the physisorbed state is shown with the transition state. Among all four chemisorptions shown in Figure 2, the lowest energy state is chosen for each of the above three states. The energies of physisorbed (phys), chemisorbed (chem) and transition states shown above are relative to the separated system.

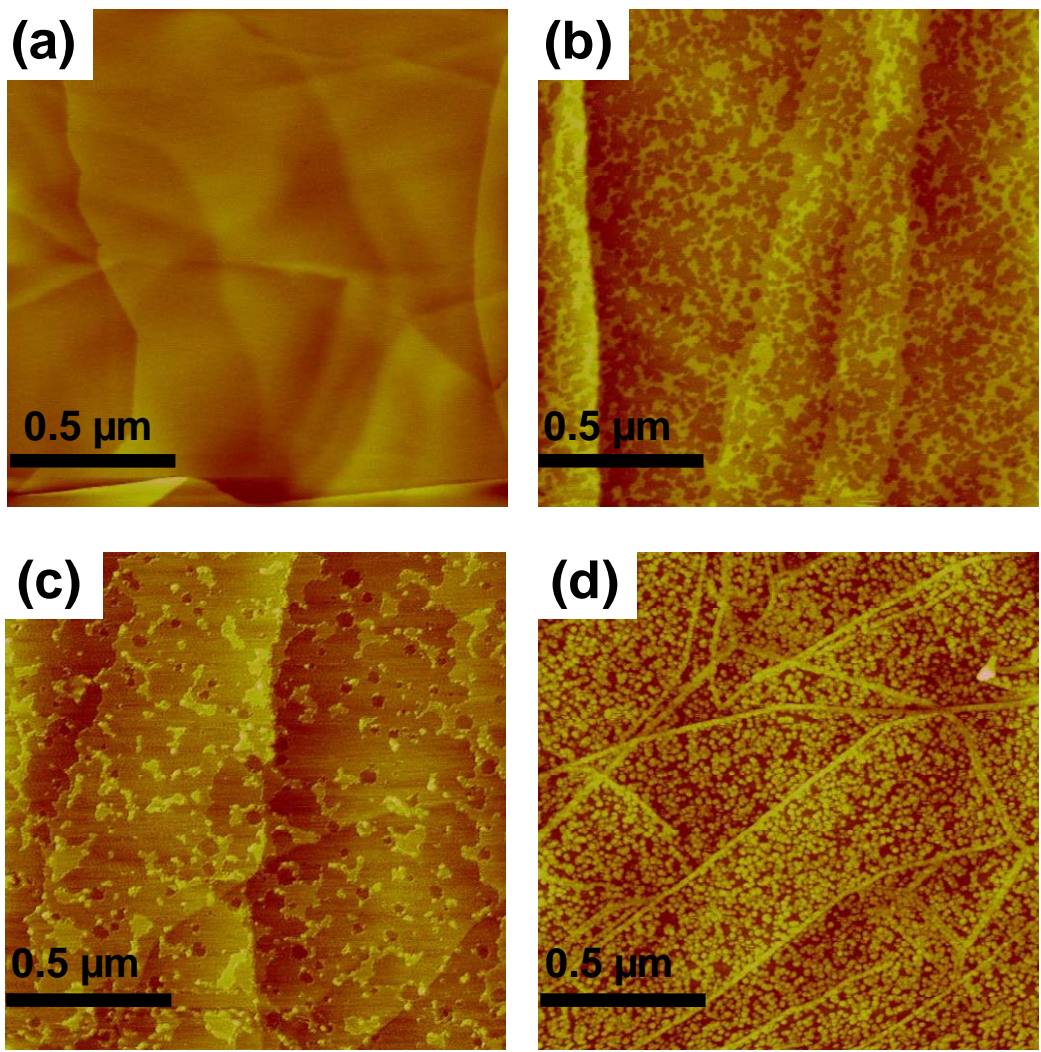

**Figure 4.** AFM images of graphite after (a) 10, (b) 30, and (c) 60 seconds ozone exposure at 500 K. (b) and (c) show ozone exposure induced etching of the graphite surface, but (a) does not show any etching. (d) ALD growth of alumina by 20 cycles of TMA/H$_2$O on sample (a) graphite with 10 seconds ozone exposure. Uniformly nucleated alumina films are visible in the AFM image.

**Supplementary Material**

Reaction of ozone with graphene at the equilibrium

We estimate physisorbed coverage of $O_3$ by using Langmuir Isotherm model [1]. The ozone flux arriving on graphene is given by $p[O_3]/\sqrt{2\pi m k_B T}$ from the ideal gas model, where $p[O_3]$ and $m$ are ozone partial pressure and mass, respectively. The number of ozone molecules to be physisorbed in an area $A$ per unit time is given by

$$\frac{p[O_3]}{\sqrt{2\pi m k_B T}} A S_{O3}(1-\theta). \qquad (1)$$

Here, $S_{O3}$ is the sticking coefficient, and $\theta$ is the physisorption coverage. Meanwhile, the desorption rate is given by $v_0 \exp(-E_b^{O3}/k_B T)$, where $v_0$ is the attempt frequency and $E_b^{O3}$ is the physisorption binding energy. The number of desorbed ozone molecules in an area $A$ per unit time is given by

$$\frac{A}{\sigma}\theta v_0 \exp(-E_b^{O3}/k_B T), \qquad (2)$$

where $\sigma$ means the cross section of ozone.

From equations (1) and (2), we can get the physisorption coverage at the equilibrium as a function of $p[O_3]$, $S$, $T$.

$$\theta^{-1} = 1 + \frac{\sqrt{2\pi m k_B T} v_0 \exp(-E_b^{O3}/k_B T)}{p[O_3] S_{O3} \sigma}. \qquad (3)$$

We use $v_0 = 2\times 10^{12}\ \text{sec}^{-1}$ and $E_b^{O3} = 0.25$ eV from our density functional calculation, and $\sigma = 10^{-19}\ \text{m}^2$. Figure S1 shows ozone coverage as a function of ozone partial pressure for different sticking coefficients $S_{O3}$ at T=500 K ($k_B T$=0.042 eV). By using $S_{O3}$=4.5×10$^{-5}$ as reported for $NO_2$ and CNT [2], we estimate θ~10$^{-7}$ ML for the ozone partial pressure 0.01 MPa used in our experiment. From our calculated physisorption binding energy of $O_2$ ($E_b^{O2} = 0.15$ eV), we get a similar result as shown in Figure S1. Since these two physisorption coverages are too small to effectively functionalize our HOPG, we consider chemisorptions as will be shown below.

We consider two chemical reactions occurring between ozone and graphene. One is dissociative chemisorption of ozone leading to epoxide and $O_2$, and the other is reverse reaction forming ozone from epoxide and $O_2$. Based on the Arrhenius equation, the chemisorption rate of physisorbed $O_3$ is given by

$$u = v_0 \exp(-\Delta/k_B T). \qquad (4)$$

Δ=0.72 eV is the energy barrier for the dissociative chemisorption and we use the attempt frequency at the physisorption state $v_0 = 2\times 10^{12}\ \text{sec}^{-1}$. The desorption rate for the reverse reaction is given by

$$v = v_0' \exp(-\Delta'/k_B T). \qquad (5)$$

We get $v_0' = 2 \times 10^{13}$ sec$^{-1}$ for the attempt frequency of epoxide and $\Delta' = 0.80$ eV for the reverse reaction energy barrier from our calculations. All of physisorbed O$_3$ molecules have a chance to be chemisorbed, so the number of chemisorption events per unit area and time is $\theta_{O3} \times u$, where $\theta_{O3}$ is the O$_3$ physisorption coverage. But, not all of epoxides are accessible for the reverse reaction, instead they need physisorbed O$_2$ molecules nearby. When we use $\theta_{O2}$ for the O$_2$ physisorption coverage, $\theta_{O2}$ fraction of epoxides have a chance to be desorbed. This allows us to write the reverse reaction events per unit area and time as $\theta_O \times \theta_{O2} \times v$, where $\theta_O$ is the epoxide coverage. At the equilibrium where the chemisorption and the reverse reaction rates are the same, the following equation holds.

$$\theta_{O3} \times u = \theta_O \times \theta_{O2} \times v \qquad (6)$$

By using equations (4) and (5), we have

$$\theta_O = \frac{\theta_{O3}}{\theta_{O2}} \frac{v_0 \exp(-\Delta/k_B T)}{v_0' \exp(-\Delta'/k_B T)}. \qquad (7)$$

For the physisorption coverages of ozone ($\theta_{O3}$) and O$_2$ ($\theta_{O2}$), we use the previously obtained formula equation (3). In Figure S1, we have seen that the physisorption coverage was very low, so we can safely keep only the second term in equation (3) for the moderate pressure as in our experiment. We can simplify equation (7) in terms of the pressure and the sticking coefficient ratios of ozone and O$_2$ as shown in the following equation.

$$\theta_O \cong \sqrt{\frac{2}{3} \frac{p[O_3]}{p[O_2]} \frac{S_{O3}}{S_{O2}} \frac{\exp(-E_b^{O2}/k_B T)}{\exp(-E_b^{O3}/k_B T)} \frac{v_0 \exp(-\Delta/k_B T)}{v_0' \exp(-\Delta'/k_B T)}}$$

We assumed that ozone and O$_2$ have the same attempt frequency $v_0$ at their physisorbed states and also they have the same cross section ($\sigma$), but they are not a significant factor in our result. Although we may use $p[O_3]=0.1\, p[O_2]$ as guessed from our experimental facility, we can plot the epoxide coverage as a function of temperature for different ratio of r = $p[O_3]S_{O3}/p[O_2]S_{O2}$ by using our calculated binding energies ($E_b^{O3}, E_b^{O2}$), energy barriers ($\Delta$, $\Delta'$) and attempt frequencies ($v_0$, $v_0'$). Figures S2(left) shows epoxide coverage versus temperature for different values of r. One can see that when the temperature is low enough, the epoxide coverage always reaches 1 ML after equilibrium. But as T increases, the coverage decreases exponentially. When T=500 K and $p[O_3]/p[O_2]=0.1$ as in our experiment, the estimated coverage is 0.5 ML assuming the same sticking coefficient (r=0.1). From Figure S2(left), we can suggest that the equilibrated epoxide coverage can be controlled not only by temperature, also by partial pressure of ozone.

Furthermore, we can consider the time for reaching equilibrium. For this purpose, we estimate how fast physisorbed O$_3$ chemisorb into epoxide by $\theta_{O3} \times u$, and we plot epoxide growth rate using equations (3) and (4) in Figure S2(right) by assuming $S$=4.5×10$^{-5}$ and $p[O_3]$=0.01 MPa. We can see that if temperature too low, it takes huge amount of time, for example, even for room temperature (T=300 K), it takes 10$^5$ seconds (~27 hours) to reach the equilibrium coverage 1 ML. When T=500 K as in our experiment, the equilibration time is reasonably 25 seconds.

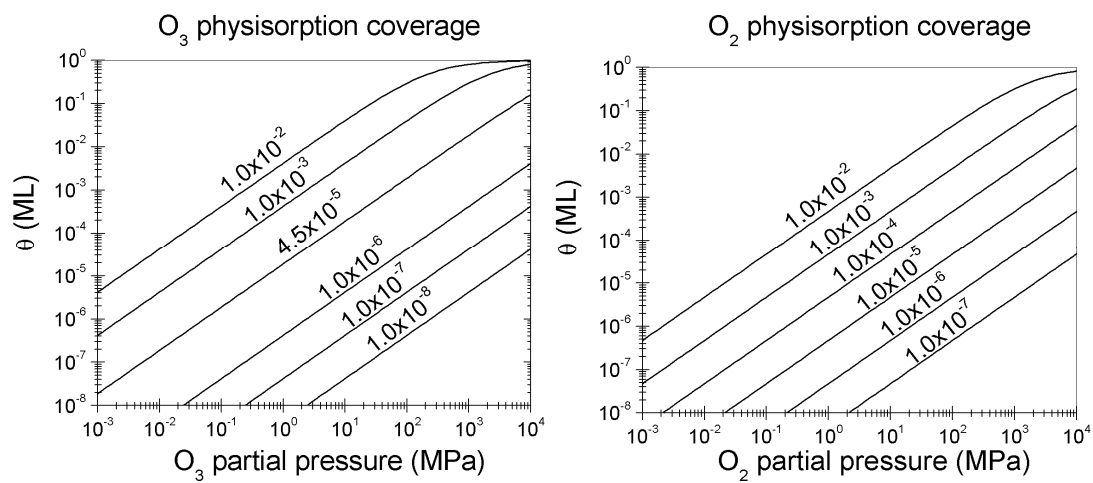

**Figure S1.** The physisorption coverage of $O_3$ (left) or $O_2$ (right) onto graphene versus its partial pressure.

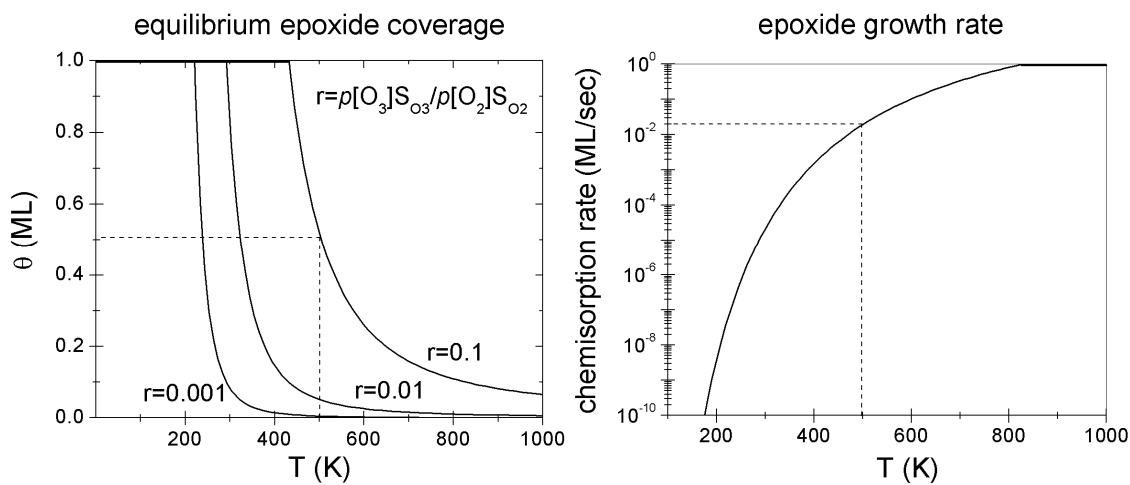

**Figure S2.** (left) The equilibrium epoxide coverage versus temperature for different values of r which is the product of partial pressure and sticking coefficient ratios between ozone and $O_2$. The dashed line indicates our experimental condition (T=500 K, $p[O_3]/p[O_2]$=0.1), which gives 0.5 ML assuming the same sticking coefficient. (right) The epoxide growth rate versus temperature by using $S$=4.5×10$^{-5}$ and $p[O_3]$=0.01 MPa. One can estimate the time for reaching equilibrium coverage shown in the left figure, where it takes about 25 seconds for T=500 K assuming r=0.1.

[1] Somorjai, G. A. *Introduction to Surface Chemistry and Catalysis*; Wiley: New York, USA, 1994; p. 303.
[2] Peng, S.; Cho, K.; Qi, P.; Dai, H. *Chem. Phys. Lett.* **2004**, *387*, 271.